# A CLASS OF MULTI-CHANNEL COSINE MODULATED IIR FILTER BANKS

V. Sundaram[*], M. J. Dehghani[#], K. M. M. Prabhu and R. Aravind

E-mail: sundaram_vanka@rediffmail.com, dehghani@sutech.ac.ir, prabhu_kmm@hotmail.com, aravind@tenet.res.in

Department of Electrical Engineering, Indian Institute of Technology - Madras, India

[*]Currently with Redpine Signals Inc., Road #2, Sagar Society, Hyderabad, India

[#]Corresponding author. IEEE member, Currently at Shiraz University of Technology, Modarres Boulevard, P.O. Box 313 Code 71555, Shiraz, Iran.

**ABSTRACT**

This paper presents a class of multi-channel cosine-modulated filter banks satisfying the perfect reconstruction (PR) property using an IIR prototype filter. By imposing a suitable structure on the polyphase filter coefficients, we show that it is possible to greatly simplify the PR condition, while preserving the causality and stability of the system. We derive closed-form expressions for the synthesis filters and also study the numerical stability of the filter bank using frame theoretic bounds. Further, we show that it is possible to implement this filter bank with much lower number of arithmetic operations when compared to FIR filter banks with comparable performance. The filter bank's modular structure also lends itself to efficient VLSI implementation.

## I. INTRODUCTION

Multirate filter banks have many applications, ranging from subband audio and video coding, multicarrier modulation systems and acoustic echo cancellation. A typical multi-channel filter bank is shown in Figure 1. The analysis filters extract the frequency components of the input signal using a suitably designed set of bandpass filters. The different subband components may be processed independently, and then combined to yield the reconstructed signal. A Perfect Reconstruction (PR) filter bank is designed such that its output is a time-delayed version of its input. PR filter banks have been well studied in literature [1, 2]. Modulated filter banks form an important subclass of filter banks. The constituent filters of a multi-channel modulated filter bank are obtained by modulating a single lowpass prototype filter. The modulating sequences can be complex exponentials (DFT filter banks) or a co-sinusoids (Cosine Modulated Filter Banks) of different frequencies. Perfect Reconstruction Cosine Modulated Filter banks (PR-CMFBs) are a natural choice for the subband analysis of real signals. Most CMFB designs proposed to date have been based on FIR prototype filters [3, 4, 5]. We note that there are distinct advantages in using FIR filters, such as ease of design and guaranteed stability. However, a major known drawback of FIR-based designs is the filter length needed for even modest design specifications.

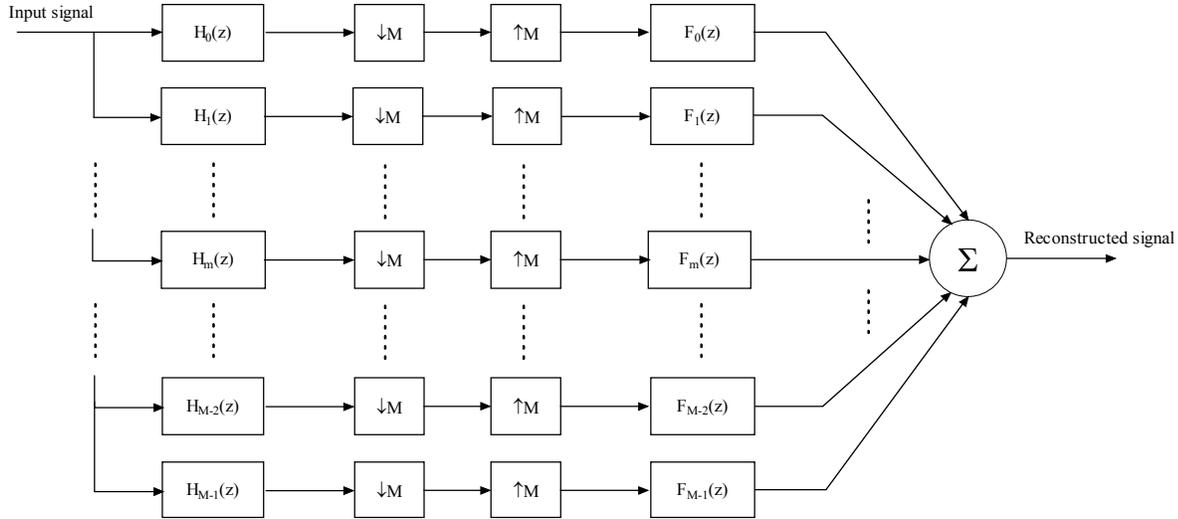

Figure 1: Multichannel filter bank

IIR filters on the other hand can offer lower system delay and give comparable performance with lower filter orders. However, these benefits come at the cost of satisfying complicated PR conditions while ensuring causality and stability of the filter bank. A general framework was for designing both FIR and IIR biorthogonal CMFBs was discussed in [6]. Other proposed IIR-CMFB include non-uniform recombination CMFBs [7], factorizations based on lifting schemes [8], and numerically optimized prototype filters [9]. In some cases, even the PR condition was relaxed to meet IIR design constraints [10]. In this paper, we present a class of causal, stable multi-channel PR-IIR-CMFBs by simplifying the PR conditions for a chosen class of IIR filters. We also show that the proposed class of filter banks require significantly lower arithmetic operations to implement when compared to comparable FIR-CMFBs. Finally, we analyze the numerical stability of the these filter banks and derive closed-form expressions for the frame bounds, and hence the frame ratio. A preliminary account of this work is reported earlier [11].

This paper is organized as follows. In Section II, we discuss some useful results from filter bank theory, and apply them to CMFBs. In Section III, we present the proposed class of IIR filters, and show that the PR condition is considerably simplified. Section V is devoted to the derivation of frame theoretic bounds for the proposed CMFB. Some design examples are presented in Section VI, and we make our concluding remarks in Section VII. We adopt the following notation: Matrices and vectors are denoted by boldface letters. For a matrix or a vector **X**, its conjugate-transpose is denoted by $\mathbf{X}^*$, and its (element-by-element) transpose by $\mathbf{X}^T$. An $M \times M$ identity matrix is denoted by $\mathbf{I}_M$, while $\mathbf{J}_M$ denotes an $M \times M$ reverse identity matrix (1's only along the main anti-diagonal and 0's elsewhere)

## II. POLYPHASE REPRESENTATION OF FILTER BANKS

### A. ANALYSIS OF MULTICHANNEL FILTER BANKS

Consider an $M$-channel PR filter bank with a downsampling factor $D \leq M$. In this paper, we restrict to the case $D = M$, which are called critically sampled filter banks (CSFB). The analysis filter set is denoted by $\{H_k(z)\}$ and the sythesis filter set $\{F_k(z)\}$, for k = 0,1,...,M-1. For efficient implementation, we use the polyphase structure for all the subband filters $H_k(z)$ and $F_k(z)$ [1]. Using the polyphase representation of the analysis and synthesis filters

$$H_k(z) = \sum_{l=0}^{M-1} H_{k,l}(z^M) z^{-l} \tag{1a}$$

and

$$F_k(z) = \sum_{l=0}^{M-1} F_{k,l}(z^M) z^{-(M-1-l)}, \quad k = 0, 1, ..., M-1, \tag{1b}$$

where $H_{k,l}(z)$ and $F_{k,l}(z)$ are the polyphase elements of $H_k(z)$ and $F_k(z)$, type I and type II, respectively, the well-known filter bank structure may be derived [1,2]. Collecting all the polyphase terms for each subband filter in (1a) and (1b), one can define the analysis polyphase matrix

$$\mathbf{A}_M(z) = \left[H_{k,l}(z)\right]_{M \times M}, \quad 0 \le k \le M-1, 0 \le l \le M-1. \tag{2a}$$

and the synthesis polyphase matrix

$$\mathbf{S}_M(z) = \left[F_{l,k}(z)\right]_{M \times M}, \quad 0 \le k \le M-1, 0 \le l \le M-1. \tag{2b}$$

The filter bank satisfies the PR condition if

$$\mathbf{S}_M(z)\mathbf{A}_M(z) = z^{-n_0}\mathbf{I}_M \tag{3}$$

where $n_0$ is delay of the filter bank.

*B. APPLICATION TO CMFB*

The general results discussed above will be stated for the CMFBs of interest. In this paper, we restrict ourselves to biorthogonal CMFBs, where the analysis filters $h_m(n)$ and synthesis filters $f_m(n)$ are derived from their respective prototype filters p(n) and q(n) according to

$$h_m(n) = 2p(n)\cos((2m+1)\pi/2M(n-(2M-1)/2)+\phi(m)), \quad m=0,1,...,M-1 \qquad (4a)$$

$$f_m(n) = 2q(n)\cos((2m+1)\pi/2M(n-(2M-1)/2)-\phi(m)), \quad m=0,1,...,M-1 \qquad (4b)$$

$\phi(m)$ can be conveniently chosen to be $(-1)^m \pi/4$ [3].

Writing the prototype filter P(z) in terms of their (type-I) polyphase components,

$$P(z) = \sum_{k=0}^{2M-1} P_k(z^{2M})z^{-k} \qquad (5)$$

$$Q(z) = \sum_{k=0}^{2M-1} Q_k(z^{2M})z^{-k} \qquad (6)$$

It was shown [5] that

$$\mathbf{A}_M(z) = \mathbf{C}_a \mathbf{P}(z) \qquad (7)$$

$$\mathbf{D}_M(z) = \mathbf{Q}(z)\mathbf{C}_s^T \qquad (8)$$

where

$$\mathbf{P}(z) = \begin{bmatrix} diag(P_0(-z^2), P_1(-z^2),...,P_{M-1}(-z^2)) \\ diag(z^{-1}P_M(-z^2), z^{-1}P_{M+1}(-z^2),...,z^{-1}P_{2M-1}(-z^2)) \end{bmatrix} \qquad (9)$$

$$\mathbf{Q}(z) = \begin{bmatrix} diag\{z^{-1}Q_{2M-1}(-z^2), z^{-1}Q_{2M-2}(-z^2),...,z^{-1}Q_M(-z^2)\} \\ diag\{Q_{M-1}(-z^2), Q_{M-1}(-z^2), ..., Q_0(-z^2)\} \end{bmatrix}^T, \qquad (10)$$

$$\mathbf{C}_a = [2\cos((2j+1)\pi/2M(k-(2M-1)/2)+\phi(j))]_{M \times 2M} \qquad (11)$$

$$\mathbf{C}_s = [2\cos((2j+1)\pi/2M(2M-1-k-(2M-1)/2)-\phi(j))]_{M \times 2M} \qquad (12)$$

where $0 \leq j \leq M-1, 0 \leq k \leq 2M-1$. $\mathbf{C}_s$ and $\mathbf{C}_a$ were also shown to satisfy

$$\mathbf{C}_s^T \mathbf{C}_a = 2M \begin{bmatrix} \mathbf{I}_M + \mathbf{J}_M & \mathbf{0} \\ \mathbf{0} & \mathbf{I}_M - \mathbf{J}_M \end{bmatrix}$$

The PR condition for this system imposes the following conditions on the prototype polyphase components $\{P_k(z)\}$ and $\{Q_k(z)\}$:

$$\begin{bmatrix} P_0 & P_M & 0 & . & & . & . & . & 0 \\ P_{M-1} & -P_{2M-1} & 0 & . & & . & . & . & 0 \\ 0 & 0 & P_1 & P_{M-1} & 0 & . & . & & 0 \\ 0 & 0 & P_{M-2} & -P_{2M-2} & 0 & . & . & & 0 \\ . & . & . & . & & . & . & & . \\ . & . & . & . & & . & . & & . \\ . & . & . & . & & . & P_{M-1} & P_{2M-1} \\ 0 & 0 & 0 & 0 & & . & . & P_0 & -P_M \end{bmatrix} \begin{bmatrix} Q_{2M-1} \\ Q_{M-1} \\ Q_{2M-2} \\ Q_{M-2} \\ . \\ . \\ Q_M \\ Q_0 \end{bmatrix} = \begin{bmatrix} Cz^{-n_0} \\ 0 \\ Cz^{-n_0} \\ 0 \\ . \\ . \\ Cz^{-n_0} \\ 0 \end{bmatrix}, \quad (13)$$

where $C$ is some real scaling factor, and $P_k \equiv P_k(-z^2)$ and $Q_k \equiv Q_k(-z^2)$, for $k = 0, 1, ..., 2M-1$. This filter bank can be implemented using the polyphase filters $P_k$ and $Q_k$ as depicted in Figures 2 and 3 respectively [1].

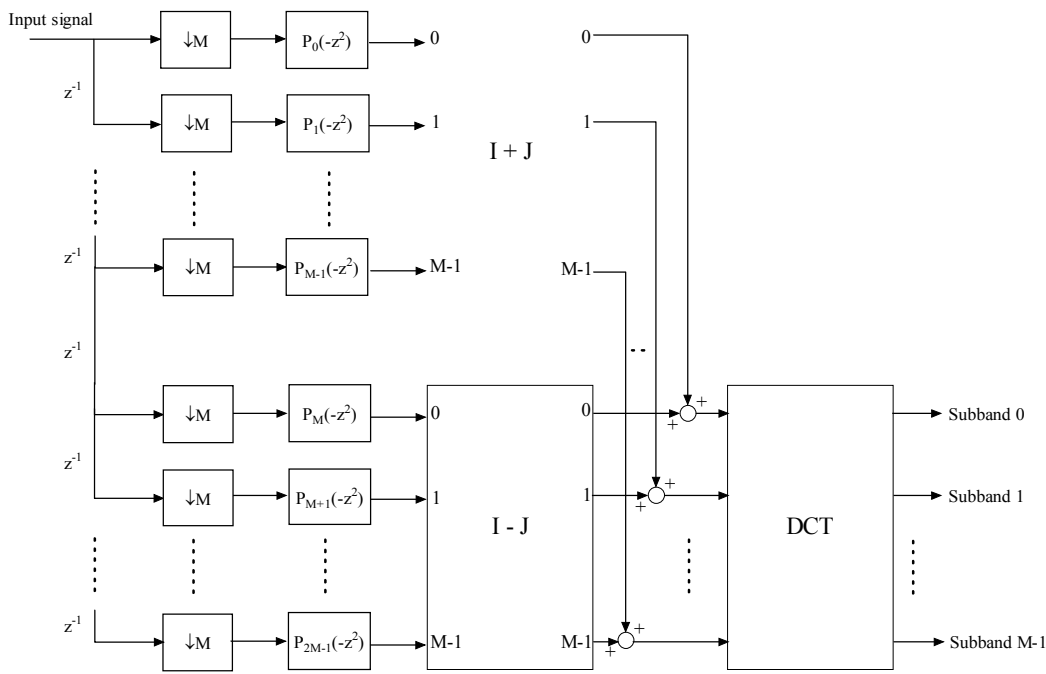

**Figure 2: Polyphase implementation of a CMFB (analysis)**

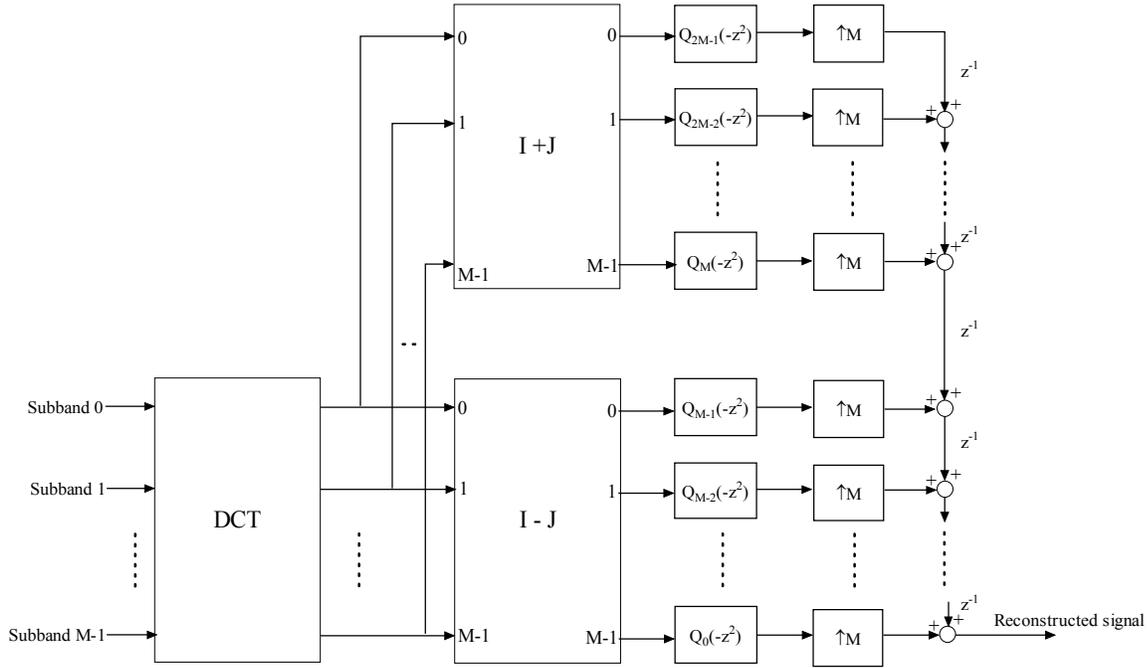

**Figure 3: Polyphase implementation of a CMFB (synthesis)**

## III. PROPOSED IIR-BASED CMFB DESIGN

The proposed *M*-channel IIR CMFB is based on designing a prototype LPF with a transfer function:

$$P(z) = \frac{\sum_{k=0}^{K} a_k z^{-k}}{1 + \sum_{m=1}^{N} b_m z^{-mD}} \quad (14)$$

where *K* and *ND* are the numerator and denominator orders, respectively. An equiripple solution to this class of filters has been given in [12, 13]. In the present work, a similar approach was used to design the prototype transfer function with $K = 2M-1$ and $D = 2M$. The design procedure places all the zeros of *P(z)* on the unit circle. This means that if $z = z_0$ is a zero of *P(z)*, so is $1/z_0$ (since $z^* = 1/z$ if $|z|=1$). This implies a symmetry among the coefficients $a_k = \pm a_{2M-1-k}$, $k = 0,1,...,M-1$.

Henceforth, the results derived here for $a_k = a_{2M-1-k}$ also hold when $a_k = -a_{2M-1-k}$.

In the following, we will show that with $D=2M$, one can design an IIR CMFB whose $2M$ polyphase components are easily derivable. Choosing $K = 2M-1$ and $D = 2M$, we have a critically sampled CMFB (CS-CMFB). The case $D < 2M$ will correspond to an oversampled CMFB. Substituting $K = 2M-1$ and $D = 2M$, in (14) gives

$$P(z) = \sum_{k=0}^{2M-1} a_k \left( \frac{1}{1 + \sum_{m=1}^{N} b_m z^{-2mM}} \right) z^{-k} \qquad (15)$$

Comparing (15) with (5) it is evident that

$$P_k(z) = a_k E(z), \qquad (16)$$

where $E(z) = \dfrac{1}{1 + \sum_{m=1}^{N} b_m z^{-m}}$, for $k=0,1,...,2M-1$.

All the $P_k(z)$'s are thus scaled versions of the same all-pole filter $E(z)$. We note that the stability of $P(z)$ guarantees that the roots of $1 + \sum_{m=1}^{N} b_m z^{-m}$ are inside the unit circle, and therefore, $\{P_k(z)\}$ are necessarily causal and stable. Substituting (16) in (13),

$$E(-z^2)\begin{bmatrix} a_0 & a_M & 0 & . & . & . & . & 0 \\ a_{M-1} & -a_{2M-1} & 0 & . & . & . & . & 0 \\ 0 & 0 & a_1 & a_{M-1} & 0 & . & . & 0 \\ 0 & 0 & a_{M-2} & -a_{2M-2} & 0 & . & . & 0 \\ . & . & . & . & . & . & . & . \\ . & . & . & . & . & . & . & . \\ . & . & . & . & . & . & a_{M-1} & a_{2M-1} \\ 0 & 0 & 0 & 0 & . & . & a_0 & -a_M \end{bmatrix} \begin{bmatrix} Q_{2M-1}(-z^2) \\ Q_{M-1}(-z^2) \\ Q_{2M-2}(-z^2) \\ Q_{M-2}(-z^2) \\ . \\ . \\ Q_M(-z^2) \\ Q_0(-z^2) \end{bmatrix} = \begin{bmatrix} Cz^{-n_0} \\ 0 \\ Cz^{-n_0} \\ 0 \\ . \\ . \\ Cz^{-n_0} \\ 0 \end{bmatrix}$$

(17)

The polyphase components $Q_k(-z^2)$ can be solved using the block nature of the matrix, as

$$Q_{2M-1-k}(-z^2) = \frac{a_k}{a_k^2 + a_{k+M-1}^2} \frac{Cz^{-n_0}}{E(-z^2)} \tag{18a}$$

$$Q_{M-1-k}(-z^2) = \frac{a_{M-k}}{a_{M-k}^2 + a_{2M-1-k}^2} \frac{Cz^{-n_0}}{E(-z^2)} \tag{18b}$$

In simplifying the above expressions, we have used the symmetry of the coefficients $a_k$, for $k=0$, 1,..., $2M$-1. Choosing $C = 1$ and $n_0 = 0$ and defining

$$s_{2M-1-k} = \frac{a_k}{a_k^2 + a_{k+M-1}^2}, \ s_{M-1-k} = \frac{a_{M-k}}{a_{M-k}^2 + a_{2M-1-k}^2}, \ k=0, 1, ..., M\text{-}1, \tag{19}$$

we get

$$Q_{2M-1-k}(-z^2) = s_{2M-1-k}/E(-z^2), \ Q_{M-1-k}(-z^2) = s_{M-1-k}/E(-z^2), \ k=0, 1, ..., M\text{-}1. \tag{20}$$

The above results also show that the analysis polyphase matrix is invertible and hence the FB allows stable reconstruction.

## IV. NUMERICAL STABILITY

Frame theory deals with overcomplete (redundant) signal representations. Basis sets are a special case of frames that have no redundancy. There has been extensive work on the frame theoretic interpretation of filter banks [14, 15, 16]. Results from frame theory can be used to study the numerical properties of a filter bank. A useful property of a filter bank is its numerical stability in the Bounded-Input-Bounded-Output (BIBO) sense. Well-designed filter banks capture all frequency components of a (finite energy) input signal well, while maintaining good subchannelization, implying high numerical stability. This is reflected in the total subband energy being "close" to the input signal energy.

Frame theoretic study of numerical stability involves the computation of a parameter known as the frame ratio. Filter banks with good frame ratios are numerically stable, and oversampled filter banks usually have better frame ratios [15].

A spanning set $\{\psi_k\}$ is said to be a frame in a signal space of square summable functions if all vectors $x$ in the space satisfy

$$A\|x\|^2 \leq \|\langle \psi_k, x \rangle\|^2 \leq B\|x\|^2,$$

where $A$ and $B$ are constants satisfying $0 < A$, and $B < \infty$. The constants $A$ and $B$ as defined as the frame bounds, yielding the frame ratio $\gamma = B/A$. For a filter bank with an analysis polyphase matrix $\mathbf{A}(\omega)$ the frame ratio is computed from the eigen values of $\mathbf{A}^*(\omega)\mathbf{A}(\omega)$ [15].

From equation (7),

$$\mathbf{A}_M^*(\omega)\mathbf{A}_M(\omega) = \mathbf{P}^*(\omega)\mathbf{C}_a^T\mathbf{C}_a\mathbf{P}(\omega) \tag{21}$$

In this expression, $\mathbf{C}_a^T\mathbf{C}_a$ can written as [1]

$$\mathbf{C}_a^T\mathbf{C}_a = 2M \begin{bmatrix} \mathbf{I}_M + \mathbf{J}_M & 0 \\ 0 & \mathbf{I}_M - \mathbf{J}_M \end{bmatrix} \tag{22}$$

Combining (9), (21) and (22)

$$\mathbf{A}_M^*(\omega)\mathbf{A}_M(\omega)$$

$$= |E(e^{j2\omega})|^2 \begin{bmatrix} a_0^2+a_M^2 & 0 & \cdots & 0 & a_0a_{M-1}-a_Ma_{2M-1} \\ 0 & a_1^2+a_{M+1}^2 & \cdots & a_1a_{M-2}-a_{M+1}a_{2M-2} & 0 \\ . & 0 & \cdots & 0 & . \\ . & a_ka_{M-1-k}-a_{k+M}a_{2M-1-k} & \cdots & a_k^2+a_{k+M}^2 & . \\ a_{M-1}a_0-a_{2M-1}a_M & 0 & \cdots & 0 & a_{M-1}^2+a_{2M-1}^2 \end{bmatrix}$$

(23)

Further simplification results by noting that

$$a_k a_{M-1-k} - a_{k+M} a_{2M-1-k} = a_k a_{M-1-k} - a_{2M-1-(M-1-k)} a_{2M-1-k} = a_k a_{M-1-k} - a_k a_{M-1-k} = 0 \quad (24)$$

This would leave the matrix in equation (23) in diagonal form, whose entries are all positive. So the eigenvalues are nothing but the entries along the principal diagonal. Since $E(z)$ does not have any zeros on the unit circle (the IIR prototype is stable), the only way any of the eigenvalues can go to zero is by having $a_k^2 + a_{k+M}^2 = 0$. This can be definitely avoided by proper design. As discussed earlier, the numerical properties of a system are determined by the ratio of the maximum eigenvalue, to the minimum eigenvalue of the above matrix. Evaluating them on the unit circle, we find that

$$\frac{\lambda_{\max}}{\lambda_{\min}} = \frac{\max_{k,0\leq k\leq \pi}((a_k^2+a_{k+M}^2)|E(e^{j2\omega})|^2)}{\min_{k,0\leq \omega\leq \pi}((a_k^2+a_{k+M}^2)|E(e^{j2\omega})|^2)}. \tag{33}$$

It will be seen from the design examples that the frame ratio is within acceptable limits.

The proposed structure yields considerable implementation benefits, in terms of the number of arithmetic operations required to implement the analysis and synthesis filters. Using the relation of **A** with the DCT matrix [1], along with (16) and the noble identities we arrive at the structure shown in Fig. 4. Similarly, we can obtain the structure of Fig. 5, which shows the synthesis part of the proposed CMFB. One can see that the complexity is greatly reduced, because only one filter each is to be implemented on either side of the analysis and the synthesis filter banks.

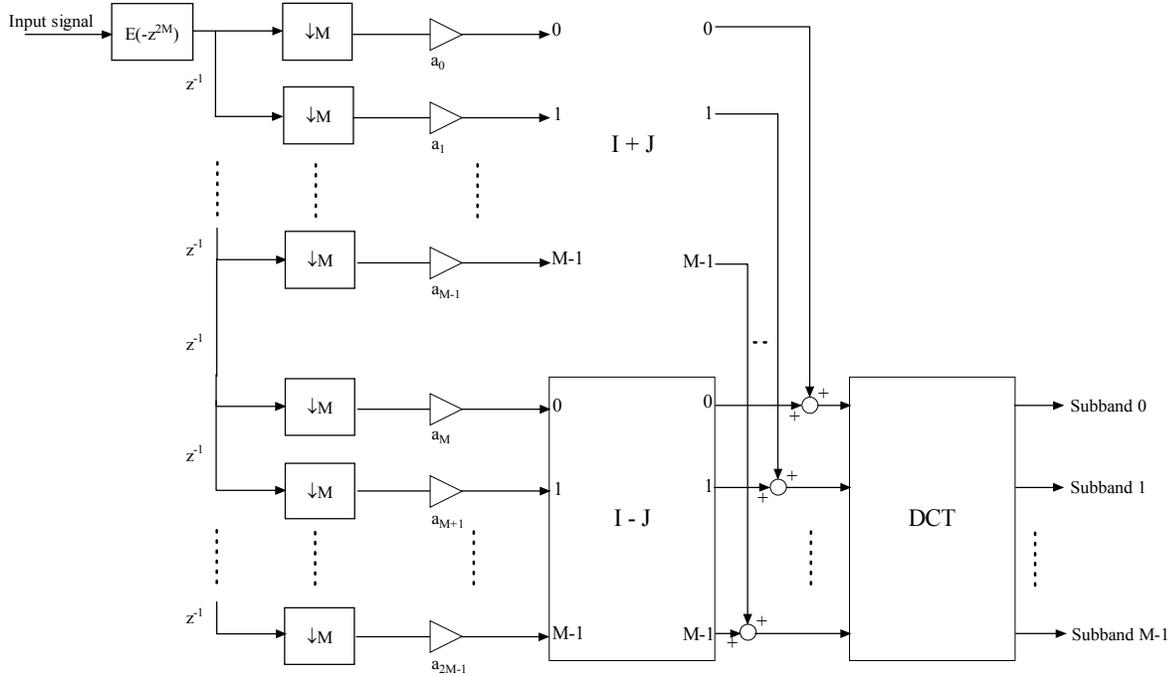

**Figure 4: Proposed IIR-CMFB (analysis)**

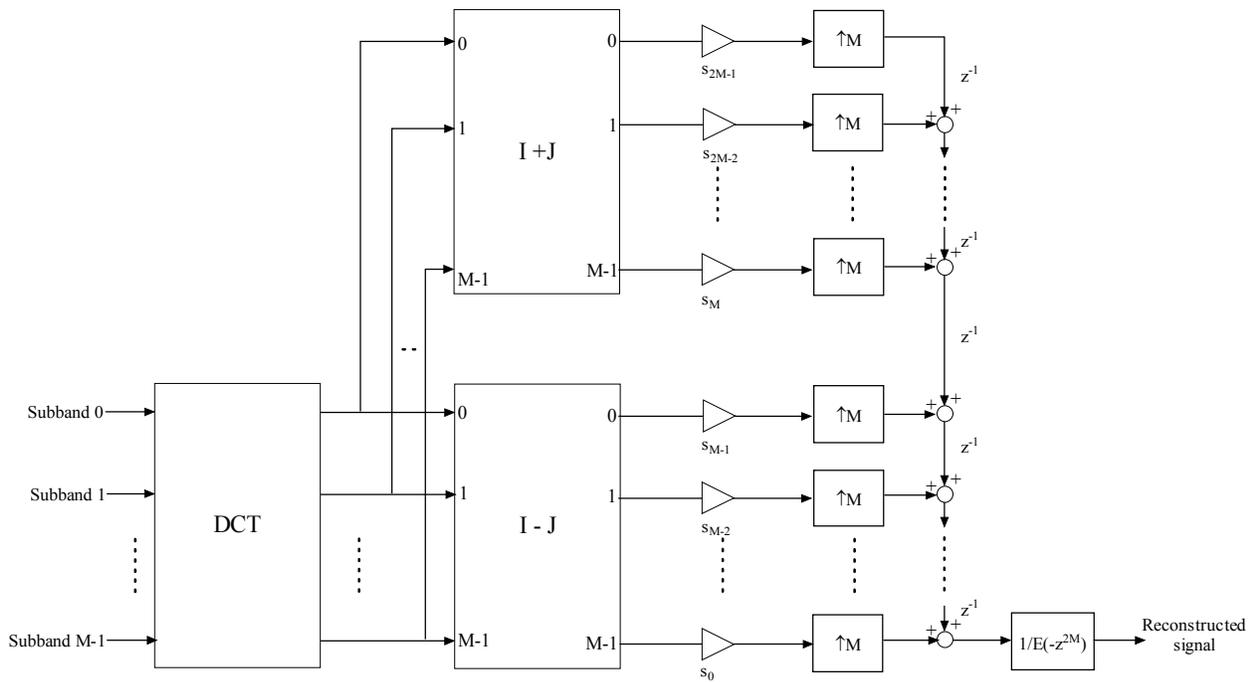

**Figure 5: Proposed IIR-CMFB (synthesis)**

The number of non-trivial multiplications (i.e., not using **I** or **J** matrices) on the analysis side for an *M*-channel case is $N+2M+M^2$. This expression was arrived at by adding the number of multiplications required to implement the filter $E(-z^{2M})$ (*N* non-zero coefficients), those required before the **I+J** and **I-J** blocks, and finally the DCT matrix, *without* fast DCT algorithms. If we make use of the fast DCT algorithms e.g., [17], we can drastically reduce the computational complexity further. Similarly, on the synthesis side, we have the same number of non-trivial multiplications equal to $N+2M+M^2$, without the use of any fast DCT algorithms.

## VI. DESIGN EXAMPLES

In [12, 13], an equiripple solution for designing filters with system function shown in (14) is presented. This algorithm provides a solution by working iteratively with magnitude of numerator and the denominator. It works as follows: given a passband ripple ($\delta_p$), normalized cut-off frequency ($f_n$), *K*, *N* and *D*, the algorithm chooses the value of the stopband attenuation ($\delta_s$), such that the resulting filter has equiripple behavior in both stopband and passband. It is shown that this algorithm is optimized in the Chebyshev sense.

We have designed the analysis prototype LPF, *p(n)*, shown in (14) based on the approach given in [12, 13]. Three cases were investigated, $M = 3,4,6$. The required filter parameters were $\delta_p = 0.05\ dB$ and normalized cut-off frequency $f_n = 1/(4M)$ for both cases. The frequency responses of analysis filters are shown in Figs. 6 and 7 respectively. It can be seen that the achievable stopband attenuation is about 31, 32 and 35 dB for *M* = 3, 4 and 6 respectively. We have seen that, to get higher stopband attenuation, we have to increase the number of zeros (*K*). In case we add more zeros more than *2M-1*, the resulting polyphase components, $P_k(z)$, in (23) cannot be all-pole filters. Consequently, the corresponding polyphase synthesis part will not be a FIR filter. In this case one should take care of stability of the synthesis part. In spite of this, the attenuation achieved is comparable to other IIR CMFBs [9] and FIR CMFBs [4]. The value of $\gamma$ has been computed in each of the cases. Tables 1, 2 and 3 show the results for $M = 3$, 4 and 6 respectively. The respective frame ratios were calculated to be 3.122, 3.57 and 4.52 dB, evidencing a high degree of numerical stability. This ensures that using the proposed design ensures numerically stable reconstruction.

## VII. CONCLUSIONS

We have discussed a class of causal, stable IIR cosine-modulated filter banks satisfying the PR condition. We have shown that imposing a suitable structure on the prototype filter greatly simplifies the PR condition, while guaranteeing both causality and stability. For a critically sampled filter bank, it was shown that this significantly reduced its implementation complexity. The numerical stability of this system was analyzed, and closed form expressions for the frame ratio was obtained. The proposed filter bank system is shown to have good frame ratios, as evident in the design examples presented. One possible way to increase the frequency selectivity of this system seems to be the use of an oversampled filter bank. This possibility is currently being investigated.

**REFERENCES**

[1] P.P.Vaidyanathan, Multirate Systems and Filter Banks, Englewood Cliffs, NJ: Prentice Hall, 1993.

[2] R. E. Crochiere and L. R. Rabiner, Chapter 7, "Multirate Digital Signal Processing", Englewood Cliffs, NJ, Prentice-Hall, Chapter 7, 1983.

[3] R. D. Koilpillai and P. P. Vaidyanathan , "Cosine-modulated FIR filter banks satisfying perfect reconstruction", IEEE Transactions on Signal Processing, Vol. 40, no. 4, pp 770-783, April 1992.

[4] J. Kliewer and A. Mertins, " Oversampled Cosine-Modulated Filter Banks with Arbitrary System Delay ", IEEE Transactions on Signal Processing, vol. 46, no. 4, pp. 941-955, April 1998.

[5] Heller, P.N., Karp, T., and Nguyen, T.Q., "A general formulation of modulated filter banks", IEEE Trans. Signal Processing, vol. 47, issue 4, pp 986-1002, April 1999.

[6] Argenti, F., "Design of biorthogonal M-channel cosine-modulated FIR/IIR filter banks", IEEE Transactions on Signal Processing, vol. 48, issue 3, March 2000.

[7] Chan, S.C., and Yin, S.S., "On the theory and design of a class of PR causal-stable IIR non-uniform recombination cosine modulated filter banks", Proc. IEEE ISCAS 2005, vol. 2, pp 1094-1097, May 2005.

[8] Chan, S.C. Mao, J.S. Yiu, P.M. Ho, K.L., "The factorization of M-channel FIR and IIR cosine-modulated filter banks and their multiplier-less realizations using SOPOT coefficients", Proc. IEEE MWSCAS 2004, vol. 2, pp 109-112, July 2004.

[9] J. S. Mao, S. C. Chan, and K. L. Ho , "Theory and Design of a Class of M-Channel IIR Cosine-Modulated Filter Banks" IEEE Signal Processing Letters, vol. 7, issue 2, pp 38-40, February 2000.

[10] Svensson, L. Lowenborg, P. Johansson, H., "A Class of Cosine-Modulated Causal IIR Filter Banks", 9th International Conference on Electronics, Circuits and Systems, 2002, vol. 3, pp 915-918.

[11] Vanka S., Dehghani M. J., Aravind R., Prabhu K. M. M., "A Class of M-Channel Reduced Complexity IIR Cosine-Modulated Filter Banks", Proc. IEEE TENCON 2003, vol 3, pp 1040-1043.

[12] H.G. Martinez and T.W.Parks, "Design of Recursive Digital Filters with Optimum Magnitude and Attenuation Poles on the Unit Circle", IEEE Trans. Acoust., Speech, Signal Processing, vol. ASSP-26, No.2, pp.150-157, Apr. 1978.

[13] H.G. Martinez and T.W. Parks, "A Class of Infinite-Duration Impulse Response Digital Filter Sampling Rate Reduction", IEEE Trans. Acoust., Speech, Signal Processing,Vol. ASSP-27, No. 4, pp. 154-162, Apr. 1979.

[14] Cvetkovic, Z. Vetterli, M., "Oversampled Filter Banks", IEEE Trans. Signal Processing, vol. 46, issue 5, pp 1245-1255, May 1998.

[15] Bolcskei, H. Hlawatsch, F. Feichtinger, H.G., "Frame-theoretic analysis of oversampled filter banks", IEEE Trans. Signal Processing, vol. 46, issue 12, pp 3256-3268, December 1998.

[16] Kovacevic, J. Dragotti, P.L. Goyal, V.K., "Filter bank frame expansions with erasures", IEEE Trans. Information Theory, vol. 48, issue 6, pp 1439-1450, June 2002.

[17] P. Yip and K.R. Rao, Fast discrete transforms: Handbook of Digital Signal Processing, Academic Press, San Diego, CA, 1987.

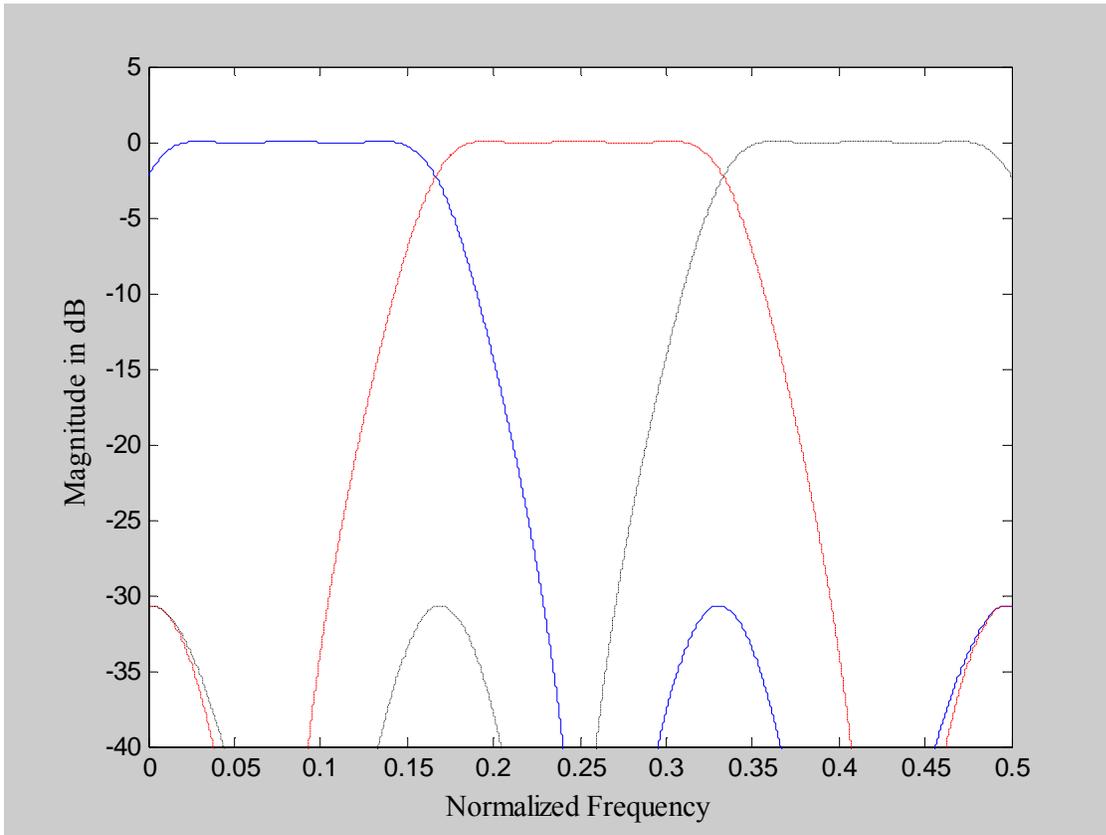

**Figure 6: Magnitude responses of analysis filters of a 3-channel CMFB**

| $a_k$ | 1 | -0.69195 | 1.02372 | 1.02372 | -0.69195 | 1 |
|---|---|---|---|---|---|---|
| $s_k$ | 0.48828 | -0.72259 | 0.49986 | 0.49986 | -0.72259 | 0.48828 |

**Table 1. Coefficients of Analysis and Synthesis Polyphase Filters for the proposed CMFB, when $M = 3$. The frame ratio was calculated to be 3.122 dB.**

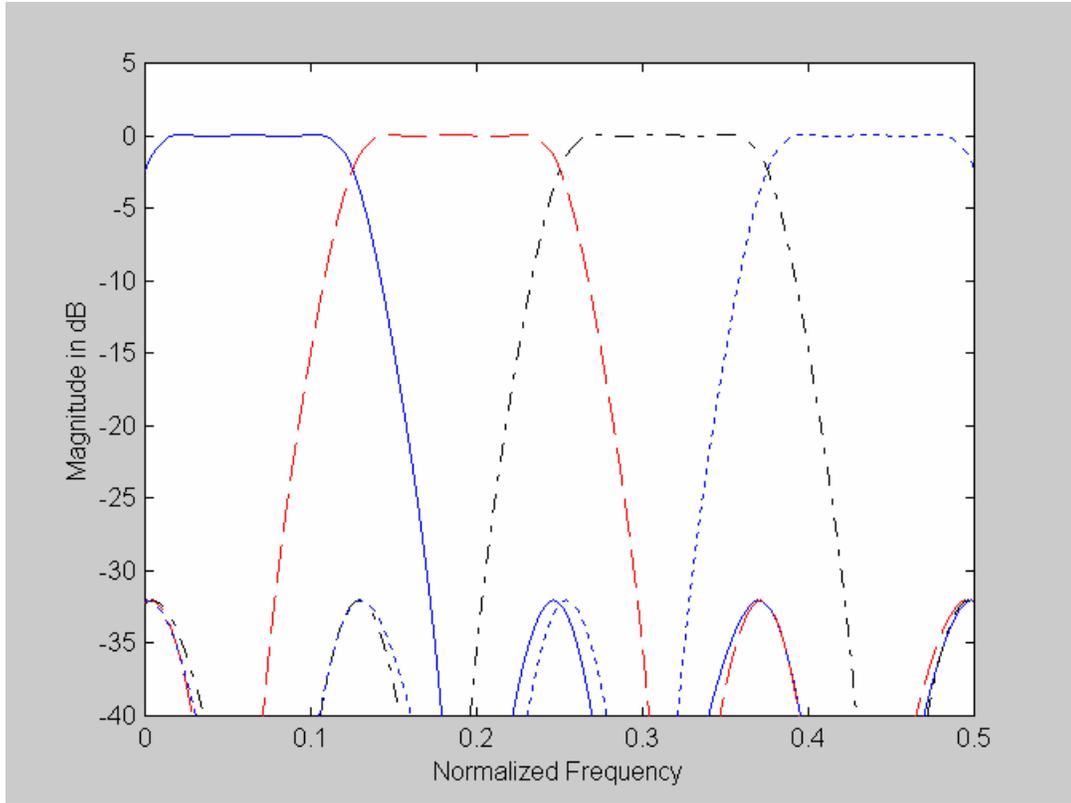

**Figure 7**: **Magnitude responses of analysis filters of a 4-channel CMFB**

| $a_k$ | 1 | 0.8720 | 1.0820 | 1.2103 | 1.2103 | 1.0820 | 0.8720 | 1 |
| --- | --- | --- | --- | --- | --- | --- | --- | --- |
| $s_k$ | 0.4057 | 0.4516 | 0.5603 | 0.4910 | 0.4910 | 0.5603 | 0.4516 | 0.4057 |

**Table 2. Coefficients of Analysis and Synthesis Polyphase Filters for the proposed CMFB, when $M$ = 4. The frame ratio was calculated to be 3.576 dB.**

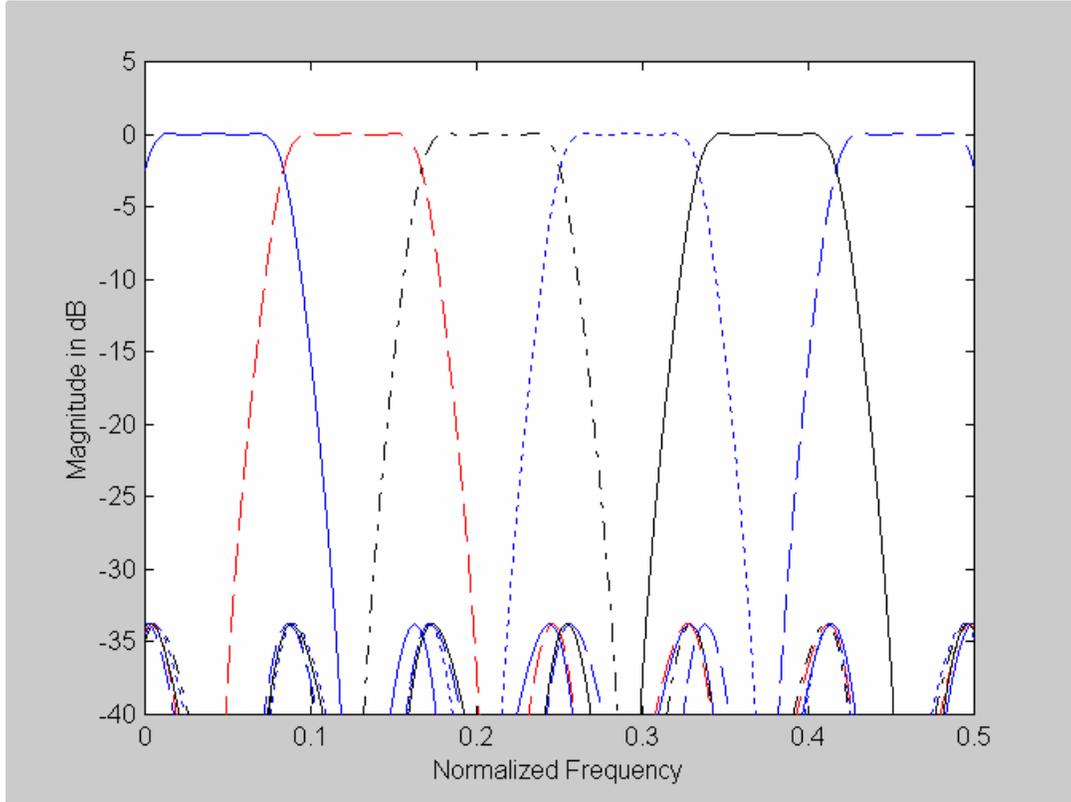

**Figure 8: Magnitude responses of analysis filters of a 6-channel CMFB**

| $a_k$ | 1 | 0.6306 | 0.7423 | 0.8623 | 0.9526 | 1 | 1 | 0.9526 | 0.8623 | 0.7412 | 0.6306 | 1 |
|---|---|---|---|---|---|---|---|---|---|---|---|---|
| $s_k$ | 0.4996 | 0.4832 | 0.5732 | 0.6668 | 0.7298 | 0.4999 | 0.4999 | 0.7298 | 0.6668 | 0.5732 | 0.4832 | 0.4996 |

**Table 3. Coefficients of Analysis and Synthesis Polyphase Filters for the proposed CMFB, when $M = 6$. The frame ratio was calculated to be 4.526 dB.**